\documentclass[12pt]{article} 

\usepackage{graphicx}

\newtheorem{example}{Example}

\begin{document}	

\title{Fully distributed and fault tolerant task management based on diffusions}

\author{Alain Bui$^\times$ \and Olivier Flauzac$^+$ \and Cyril Rabat$^+$\\[3pt]
\small $^\times$ Laboratoire PRiSM (UMR CNRS 8144)\\
\small Universit\'e de Versailles-St-Quentin-en-Yvelines\\
\small 45, avenue des Etats-Unis -- F-78035 Versailles Cedex -- France\\
\small \texttt{alain.bui@prism.uvsq.fr}\\[6pt]
\small $^+$ Laboratoire CReSTIC\\
\small Universit\'e de Reims Champagne-Ardenne,\\
\small BP 1039, F-51687 Reims Cedex 2  -- France\\
\small \texttt{$\{$olivier.flauzac,cyril.rabat$\}$@univ-reims.fr}}
\date{}

\maketitle


\begin{abstract}
The task management is a critical component for the computational grids. The aim is to assign tasks on nodes according to a global scheduling policy and a view of local resources of nodes. A peer-to-peer approach for the task management involves a better scalability for the grid and a higher fault tolerance. But some mechanisms have to be proposed to avoid the computation of replicated tasks that can reduce the efficiency and increase the load of nodes. In the same way, these mechanisms have to limit the number of exchanged messages to avoid the overload of the network.

In~\cite{BuFR07}, we have proposed two methods for the task management called active and passive. These methods are based on a random walk: they are fully distributed and fault tolerant. Each node owns a local tasks states set updated thanks to a random walk and each node is in charge of the local assignment. Here, we propose three methods to improve the efficiency of the active method. These new methods are based on a circulating word. The nodes local tasks states sets are updated thanks to periodical diffusions along trees built from the circulating word. Particularly, we show that these methods increase the efficiency of the active method: they produce less replicated tasks. These three methods are also fully distributed and fault tolerant. On the other way, the circulating word can be exploited for other applications like the resources management or the nodes synchronization. 
\end{abstract}


\section{Introduction}\label{SECTION_INTRODUCTION}

When a problem is submitted in a grid, it can be divided in a set of tasks. These tasks have to be assigned to nodes of the grid according to the resources needed for their computation and the resources owned and available on nodes. A grid has to manage a lot of resources as the storage space and the computational power, specific data, shared applications or tools. Each resource has to be identified for the global scheduling of tasks.

In \emph{Nimrod}~\cite{BuAG00}, the resources are gathered on an agent. So, the scheduling can be achieved with the agent knowledge and tasks are assigned by the agent. In Middleware \emph{NetSolve}~\cite{AAB+02}, the computational resources give an estimation of the task computation length in function of the task parameters. It lets the agent schedules the tasks according to the servers load. In such applications, the central agent is a critical point in the grid. An overload or a failure of this node can involve a grid failure or a low global efficiency. To improve the scalability, the global knowledge can be divided or shared in several servers or in a hierarchy of servers as in \emph{Globus}~\cite{Fost06} or in \emph{DIET}~\cite{ABB+06}.

Since several years, the peer-to-peer approach has been proposed for increasing the scalability and the fault tolerance of the applications. The centralization is avoided and the topology of the application is flexible: the deployment is also simplified. The authors of \emph{CONFIIT}~\cite{FlKF03} choose to let the nodes in charge of the tasks selection. The tasks parameters are sent to all nodes so they can choose to compute a task according to their local resources. But the nodes of \emph{CONFIIT} are setup in a virtual ring that has to be maintained. It involves a lot of control messages especially if the network is highly volatile and, in the worst case, the ring cannot be maintained.\\

To avoid the maintenance of a virtual structure, we proposed fully distributed methods based on a random walk. In~\cite{BuFR07} we propose a task management: the tasks parameters are sent to nodes and the tasks states are updated thanks to the circulation of a token. We proposed two tasks assignment methods called \emph{passive} and \emph{active}. For the passive method, the nodes wait for the token before selecting a task. The number of replicated tasks is low but some computational resources is unused. For the active method, the nodes select a task before receiving the token. It induces replicated tasks but it increases the global efficiency. In the following, we focus on the improvement of the active method. Indeed, we show in~\cite{BuFR07}, that this method has a better efficiency than the passive method. 

In this article, we propose to use another tool called the circulating word. Its aim is to collect node identities and to build and maintain spanning trees of the network. We propose to diffuse periodically the tasks states along these spanning trees in order to speed up the update of the nodes local knowledge.\\

In next section, we present the tools we used in our algorithm: the grid model, the random walks and the circulating word. In section~\ref{SECTION_TASK_MANAGEMENT}, we present the task management (task definition, efficiency). Then, we present our solutions and we describe the algorithms. We show in section~\ref{SECTION_EXPERIMENTAL_RESULTS} some experimental results. Finally, we conclude and we present our future works.  


\section{Preliminaries}\label{SECTION_PRELIMINARIES}

\noindent\textbf{A model for grid} In~\cite{RaBF06}, we propose a model composed of 5 layers to analyze grid applications. The three lower layers concern the network, the routing and the messages exchange protocols. Layer 4 represents the resources management for the grid and the last one the other grid components (scheduling, monitoring,~\ldots). The task management we expose here is for Layer 5. But a grid is built over four other layers and we have to take care of their impacts. We show that we can model a grid by a \emph{directed} graph $G =(V,E)$, where $V$ is a set of active nodes of the grid with $|V| = n$ and $E$ is the set of directed communication links. An active node is a resource or a node that uses resources (to compute task). In the following, we use the terms "resource", "node" and "active node" interchangeably. 

A communication link $(i,j)$ exists if and only if $j$ is a neighbor of $i$ in the grid, i.e. $i$ can directly send a message to $j$. Every node $i$ can distinguish all its links of communication and maintains a set of neighbors denoted $N_i$. We consider that all resources of the grid have a distinct identity (IP address, for example or a complete description with a specific language like \emph{RSL}~\cite{CFK+98}, in that case an indexation is needed to have better performances).

As $G$ is a communication graph, we assume it is strongly connected. Indeed, if the graph is not strongly connected at a time, there exists a sink subgraph $E(G)$: resources of $G\backslash E(G)$ cannot be reached from any node of $E(G)$. For a token circulation, it means that the token will stay in $E(G)$ and cannot reach nodes of $G\backslash E(G)$. With our method, we accept that the graph stays not strongly connected during a short time. If this transient state is too long, unreachable resources will be considered as disconnected.\\

\noindent\textbf{Random walk} A random walk is a sequence of nodes visited by a token that starts at $i$ and visits other resources according to the following transition rule: if the token is at $i$ at time $t$ then at time $t+1,$ it will be at one of the neighbors of $i$, chosen uniformly at random among $N_i$~(\cite{Lova93}). Similarly to deterministic distributed algorithms, the time complexity of random walk based token circulation algorithms can be viewed as the number of "steps" it takes for the algorithm to achieve the network traversal. With only one walk at a time (which is the case we deal), it is also equal to the message complexity. The cover time $C$ --- the average time to visit all nodes in the system --- and the hitting time denoted by $h_{ij}$ --- the average time to reach a node $j$ for the first time starting from a given node $i$ --- are two important values that appear in the analysis of random walk-based distributed algorithms. Both of them are on average bounded by $n^3$. There are three properties about random walks: \emph{percussion} -- an arbitrary node is visited in a finite time, \emph{coverage} -- all nodes  are visited in a finite time and \emph{meeting} -- several random walks will meet each other in a finite time.\\

\noindent\textbf{Circulating word} A circulating word is a tool used to collect data on a network. It has been introduced in~\cite{Lava90} for the detection of the execution termination of distributed algorithms. Particularly, it can be used to collect identities of visited nodes by a random walk. With a specific management, as proposed in~\cite{Flau01}, we are able to build a spanning tree rooted on the node that owns the token. This tree is perpetually updated and adapted to the topology when node failures occur. We can send data through the network along this tree. This application is used in our task management methods.


\section{Task management}\label{SECTION_TASK_MANAGEMENT}

We define a task by the tuple $\{id_T, id_E, p, s, r\}$ where $id_T$ is the task identity, $id_E$ the identity of the task emitter, $p$ the parameters of the task, $s$ the state of the task and $r$ the results of the task (if it has been computed). A task can be in three states: \emph{uncomputed}, \emph{in progress} (locally or in a distant node) and \emph{computed}. 

\begin{figure}
  \centering
  \begin{tabular}{cc}
    \begin{tabular}{cc}
      \includegraphics[scale=0.8]{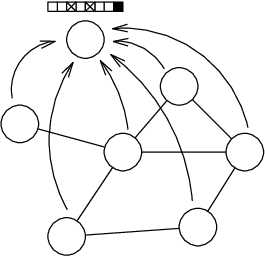} & \includegraphics[scale=0.8]{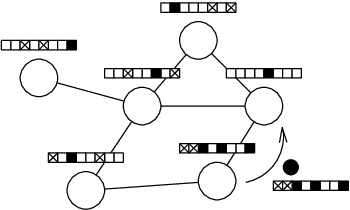} \\
      $(a)$ \vspace{0.5cm} & $(b)$
    \end{tabular} &
    \includegraphics[scale=1]{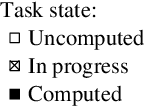}
  \end{tabular}
  \caption{Centralized task management $(a)$ and token based task management $(b)$.}\label{FIGURE_CENTRALIZED_TOKEN}
\end{figure}

With a centralized method, the tasks are gathered on a server as we present on Figure~\ref{FIGURE_CENTRALIZED_TOKEN}~$(a)$. When a task is assigned to a node, its state is modified and the task cannot be assigned to another node (excepted if a voluntary replication is achieved as in \emph{BOINC}~\cite{Ande04}). With our method, the parameters and results of the tasks are diffused to all nodes. On each node, a local set noted $\mathcal{E}$ contains a local view of the tasks states. When a node wants to compute a task, it selects an uncomputed task at random in its local set and tags it as \emph{in progress}. This new state will be updated on other nodes step by step by the token (Figure~\ref{FIGURE_CENTRALIZED_TOKEN}~$(b)$). In the same way, when a node has to submit new tasks, it add them only to its local set. The tasks will be known by other nodes when the token will visit the node and will diffuse their parameters to other nodes.

When a task is selected, its state is \emph{uncomputed} but another node can select simultaneously the same task. So, we obtain a \emph{replicated task}. This involves a waste of computational power and decreases the efficiency of the task management. To compute the efficiency of the task management, we compare the sequential execution time $t_e$ -- that is the time for one node to compute all the tasks -- and the distributed execution time $t_d$ -- that is the time for all nodes to compute all the tasks. Efficiency of the method, noted $e$, is obtained by the formula: $e=\frac{t_s}{t_e \times n} \times 100$, where $n$ is the number of nodes. We also compare the task management solutions on the number of exchanged messages.

\section{Diffusion based task management}\label{SECTION_DIFFUSION_BASED_TASKS_MANAGEMENT}

We propose three solutions based on the active method described in~\cite{BuFR07}. A token circulates at random in the network and updates the local tasks states sets of nodes. We add a circulating word in the token that is used to build spanning trees. We describe three solutions to increase the active method efficiency: method $\mathcal{D}_s$ is based on periodical diffusions, method $\mathcal{D}_f$ is based on diffusions with feedbacks and method $\mathcal{D}_m$ is based on diffusions with feedbacks followed by another diffusions.

\subsection{$\mathcal{D}_s$ method}

We define a token by a tuple $T=\{id_T, \mathcal{E}_T, W, C_T\}$ where $id_T$ is the token identity, $\mathcal{E}_T$ is the tasks states set, $W$ is a circulating word and $C_T$ is a hop counter. When a node receives a token, it increments $C_T$ and updates $W$ (by adding its identity). If $C_T$ is upper than a bound, noted $b$, the node builds from the circulating word a diffusion tree $\mathcal{T}_D$ rooted on it. Then, a diffusion of $\mathcal{E}_T$ (updated by the node set) is launched through $\mathcal{T}_D$. We define each message of the diffusion by the tuple $M_D=\{id_T, \mathcal{E}_M, \mathcal{T}_D\}$ where $id_T$ is the token identity (used to control the message validity), $\mathcal{E}_M$ is a tasks states set updated according to visited nodes and $\mathcal{T}_D$ the diffusion tree. When a message is received, the node updates its local tasks states set and forwards the message to all of its neighbors in $\mathcal{T}_D$. To reduce the messages size, $\mathcal{T}_D$ can be reduced to the subtree rooted on neighbors.\\

The frequency of diffusion launches depends on bound $b$. To reduce useless diffusions, we compute $b$ according to the current state of the global computation. Indeed, when the number of tasks to compute decreases, the local task selection method involves replicated tasks. So, we compute $b$ according to the following formula:
$$b=\min\left\{\frac{nbT}{n} * c_r, m_r \right\}$$
$nbT$ is the number of tasks to compute, $n$ is the number of nodes, $c_r$ is the refresh coefficient and $m_r$ is the minimum refresh value. When the ratio between the number of tasks and the number of nodes becomes too small, the frequency of diffusions increases: more diffusions are launched to reduce replicated tasks. To prevent the overload of the network, we specify a minimum threshold noted $m_r$. $c_r$ interacts on the diffusions frequency.

\begin{example}
Figure~\ref{FIGURE_DIFFUSION_EXAMPLE} shows an example of a single diffusion. Node 0 receives the token and here, we suppose $C_T$ becomes upper than $b$. Node 0 builds a diffusion tree from the circulating word and launches a diffusion. We show that local tasks states sets $\mathcal{E}$ are updated during the diffusion.

\begin{figure}
  \centering
  \includegraphics[scale=1]{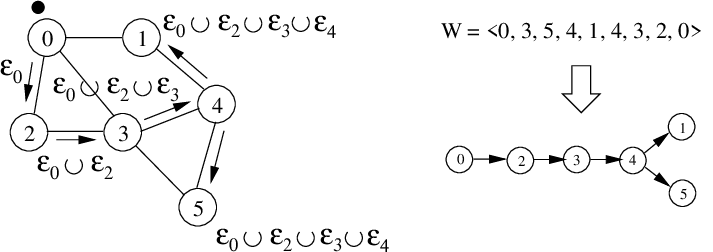}
  \caption{Example of a tasks states diffusion from a circulating word}
  \label{FIGURE_DIFFUSION_EXAMPLE}
\end{figure}
\end{example}

\subsection{$\mathcal{D}_f$ method}

On the previous example, we observe that the nodes on leaves of the diffusion tree are updated with several tasks states sets but the set of the diffusion tree root (\emph{i.e.} the initiator node) is not updated. In the same way, if a node is deeper in the tree, it is updated with several nodes sets but its local set is not diffused to other nodes. In Figure~\ref{FIGURE_DIFFUSION_EXAMPLE}, Node 5 is updated with sets of Nodes 0, 2, 3 and 4 but its tasks states set is not sent to other nodes.\\

To improve the update of local sets, we propose to add a feedback after the diffusion. Each node sends its local set to its father. At the end of the diffusion and the feedback, the initiator node receives a global view of the tasks states. To limit the number of exchanged messages, we exploit the algorithm of the distributed recursive waves described in~\cite{FlGL93}. Before sending its set to its father, each node waits for the response of each son. To support node or link failures, we add a timeout on each node reseted at each diffusion: if a node does not receive the responses of its sons before the timeout ends, it sends its set to its father. If sons responses are received later, they will be ignored (or only used to update the node).\\

In the worst case, the diffusion and the feedback take $2 \times (n-1)$ steps. If $b$ is lower than this value, several diffusions can be launched at the same time but on different diffusion trees. We need to identify each diffusion and nodes have to keep in memory their father in the diffusion and its sons. We add a diffusion counter in the token that is incremented at each diffusion. Each message of a diffusion is tagged by the counter value. On the nodes, we add two sets to keep in memory the node father and its sons corresponding to a diffusion. When a node has received a response of each son of a diffusion, it can send its own response to its father.

\subsection{$\mathcal{D}_m$ method}

After a diffusion with a feedback, the tasks states set of the initiator is updated. The nodes on the leaves of the tree are only updated by few nodes, especially if the diffusion tree has a small depth. So, after a diffusion with a feedback, we can initiate a new diffusion with the initiator set. After this diffusion, all the nodes will have the same view of the tasks states. The same diffusion tree can be used that does not cost any extra computational power for the initiator.

\subsection{Example}

\begin{figure}
  \centering
  \begin{tabular}{cc}
    \begin{tabular}{c}
      \includegraphics[scale=1]{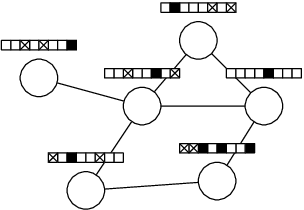} \\
      $(a)$
    \end{tabular}
    &
    \begin{tabular}{c}
      \includegraphics[scale=1]{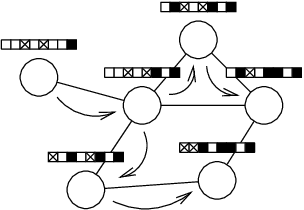} \\
      $(b)$
    \end{tabular} \\
    \begin{tabular}{c}
      \includegraphics[scale=1]{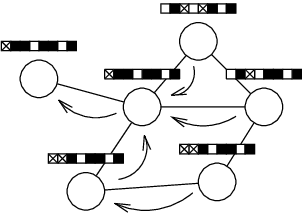} \\
      $(a)$
    \end{tabular}
    &
    \begin{tabular}{c}
      \includegraphics[scale=1]{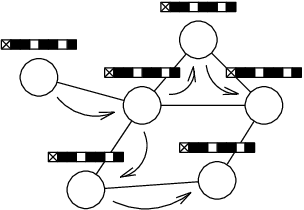} \\
      $(b)$
    \end{tabular} \\
  \end{tabular}
  \caption{Diffusion based task management: task states are updated thanks to a diffusion through a tree built from a circulating word.}\label{FIGURE_DIFFUSION}
\end{figure}

Figure~\ref{FIGURE_DIFFUSION} shows an example of an update diffusion. Figure $(a)$ presents the tasks states set of each node. Some tasks are considered as uncomputed on some nodes and are in progress on others nodes. A node receives the token and initiates a diffusion (Figure $(b)$). The tasks states sets are updated along the diffusion tree and the nodes on leaves of the tree obtain the best view. After the feedback (Figure $(c)$), the initiator receives the more recent view of all tasks. The last diffusion (Figure $(d)$) involves that all the nodes have the same view.

\section{Experimental results}\label{SECTION_EXPERIMENTAL_RESULTS}

We simulate our methods with \emph{Dasor} library~\cite{BuFR08}. We generate a set of random task lengths thanks to the log-normal law. We obtain a set of irregular task lengths. The set is sent to each node and we compute the time for the nodes to compute the tasks: here you suppose that all nodes have the same computational power. For the diffusion methods, we fix arbitrarily $c_r=1000$ and $m_r=1500$ (these values give a good compromise between the efficiency and the number of exchanged messages).\\

\begin{figure}
  \centering
  \begin{tabular}{c}
    \includegraphics[scale=0.8]{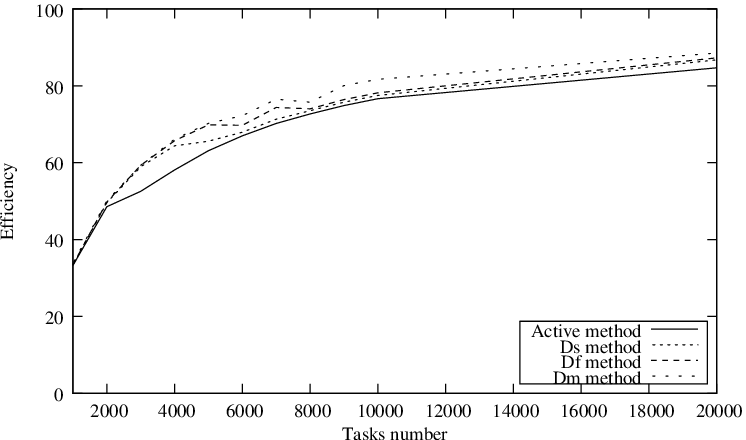} \\
    $(a)$
  \end{tabular}
  \begin{tabular}{c}
    \includegraphics[scale=0.8]{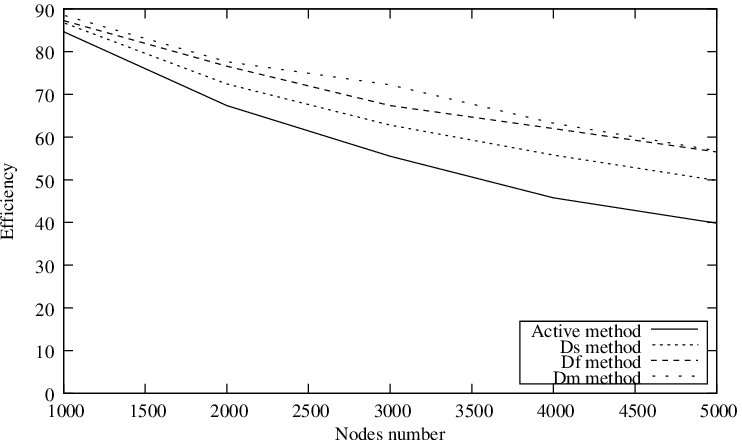} \\
    $(b)$
  \end{tabular}
  \caption{Evolution of the efficiency in function of the tasks number $(a)$ and in function of the nodes number $(b)$ .}\label{FIGURE_EXPERIMENTS1}
\end{figure}

The first series of simulations presented on Figure~\ref{FIGURE_EXPERIMENTS1}~$(a)$ shows the evolution of the efficiency in function of the grid nodes number. It presents the executions results with 1000 nodes and a number of tasks that evolves from 1000 to 20000. We remark that the diffusion methods have a better efficiency than the active method especially when the tasks number increases (about 5\% for $\mathcal{D}_m$ in average). For a small number of tasks, the efficiencies are almost identical. Indeed, at the beginning of the execution, each node selects at random a task in its local set and several nodes select the same tasks (the ratio is 1 task per node).

Figure~\ref{FIGURE_EXPERIMENTS1}~$(b)$ presents executions results with a set of 20000 tasks and a number of nodes that evolves from 1000 to 5000. When the number of nodes increases, the diffusions methods have a better efficiency (more than 15\% for $\mathcal{D}_m$). We observe that the efficiency decreases faster with the active method. The cover time of the token is higher and involves more replicated tasks: the latency between updates is higher. The diffusion methods seem to be more scalable.

\begin{figure}
  \centering
  \begin{tabular}{c}
    \includegraphics[scale=0.8]{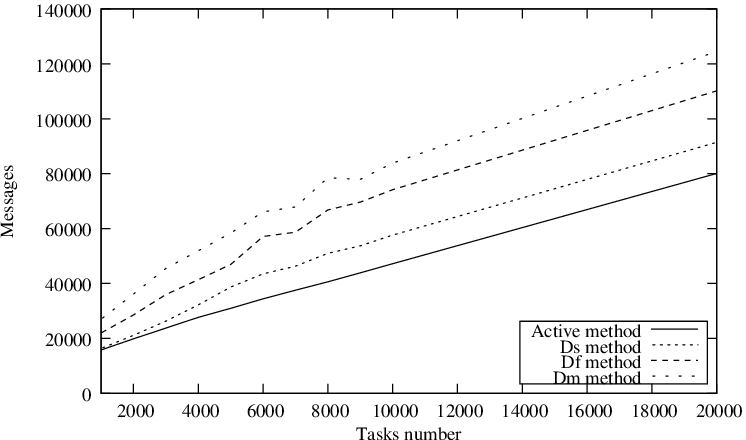} \\
    $(a)$
  \end{tabular}
  \begin{tabular}{c}
    \includegraphics[scale=0.8]{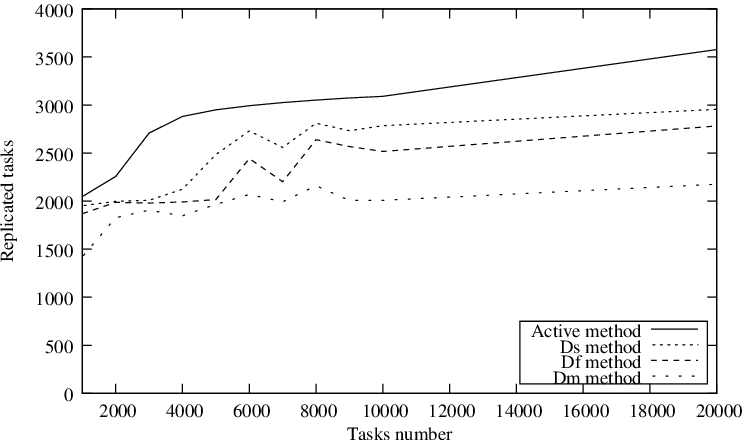} \\
    $(b)$
  \end{tabular}
  \caption{Evolution of the messages exchanges $(a)$ and of the replicated tasks $(b)$ in function of the tasks number.}\label{FIGURE_EXPERIMENTS2}
\end{figure}

On Figure~\ref{FIGURE_EXPERIMENTS2}~$(a)$, we present the number of produced messages during the computation and on Figure~\ref{FIGURE_EXPERIMENTS2}~$(b)$ the number of replicated tasks. We fix the number of nodes at 1000 and we increase the number of tasks from 1000 to 20000. We can observe that the number of messages for Method $\mathcal{D}_m$ is about twice more than the active method (with $c_r=1000$ and $m_r=1500$). About the number of replicated tasks, we have about twice less replicated tasks than the active method. The diffusion methods reduce the global load of the grid by reducing the useless computations.\\

To improve the efficiency, $c_r$ and $m_r$ can be reduced to increase the frequency of the diffusions but it induces more messages. With these experimental results, we observe that we have already a better efficiency and less replicated tasks than for the active method. We also realize others simulations series with $c_r=100$ and $m_r=100$. For 1000 nodes and 20000 tasks, it increases the efficiency by 5\% and reduce the number of replicated tasks. But it produces 4 times more messages than with the previous coefficients.


\section{Conclusion}\label{SECTION_CONCLUSION}

We propose in this article three solutions for the task management based on a random walk and a circulating word. The nodes are in charge of the local assignment of tasks according to their resources and the tasks states are updated thanks to the token. These solutions are fully distributed and are resilient to node failures. We present some experimental results and we observe a better efficiency than the active method presented in~\cite{BuFR07}. These new methods produce more messages but they reduce significantly the number of replicated tasks: it reduces the useless load of grid nodes.

These methods are based on two coefficients $c_r$ and $m_r$ that allow to modify the frequency of diffusions according to the current state of the system (number of tasks and number of nodes). We plain to automatize these parameters to take into account others grid parameters as the bandwidth of the network or the actual load of the system. Another solution is to exploit the hybrid method proposed in~\cite{BuFR07} coupled with the diffusion methods. When the ratio between the number of nodes and the number of tasks is low, we may reduce the replicated tasks.

\section*{Acknowledgments}

This work was partly supported by "Romeo"\footnote{\texttt{http://www.romeo2.fr/}}, the high performance computing center of the University of Reims Champagne-Ardenne.

\bibliographystyle{latex8}
\bibliography{biblio}
\end{document}